\documentclass[showpacs,preprintnumbers,amsmath,amssymb,preprint]{revtex4}
\usepackage{graphicx}
\usepackage{psfig}
\usepackage{bm}
\begin{document}

\def\bsig{\mbox{\boldmath $\sigma$}}                          
\def\bsig{\mbox{\boldmath $\Sigma$}}
\def\bgam{\mbox{\boldmath $\gamma$}}
\def\bgam{\mbox{\boldmath $\Gamma$}}
\def\bphi{\mbox{\boldmath $\phi$}}
\def\bphi{\mbox{\boldmath $\Phi$}}
\def\btau{\mbox{\boldmath $\tau$}}
\def\btau{\mbox{\boldmath $\Tau$}}
\def\btau{\mbox{\boldmath $\partial$}}
\def\Delc{{\Delta}_{\circ}}
\def\bp{\mid {\bf p} \mid}
\def\al{\alpha}
\def\bet{\beta}
\def\gam{\gamma}
\def\del{\delta}
\def\Del{\Delta}
\def\te{\theta}
\def\nua{{\nu}_{\alpha}}
\def\nui{{\nu}_i}
\def\nuj{{\nu}_j}
\def\nue{{\nu}_e}
\def\num{{\nu}_{\mu}}
\def\nut{{\nu}_{\tau}}
\def\2te{2{\theta}}
\def\chic#1{{\scriptscriptstyle #1}}
\def\chicl{{\chic L}}
\def\lam{\lambda}
\def\SU{SU(2)_{\chic L} \otimes U(1)_{\chic Y}}
\def\Lam{\Lambda}
\def\sig{\sigma}
\def\'#1{\ifx#1i\accent19\i\else\accent19#1\fi}
\def\O{\Omega}
\def\o{\omega}
\def\s{\sigma}
\def\D{\Delta}
\def\d{\delta}
\def\df{\rm d}
\def\8{\infty}
\def\ld{\lambda}
\def\eps{\epsilon}

\newcommand{\be}{\begin{equation}}
\newcommand{\ee}{\end{equation}}
\newcommand{\ba}{\begin{array}}
\newcommand{\ea}{\end{array}}
\newcommand{\dis}{\displaystyle}
\newcommand{\alfad}{\frac{\dis \bar \alpha_s}{\dis \pi}}
\newcommand{\bra}{\mbox{$<$}}
\newcommand{\ket}{\mbox{$>$}}

\title{Rare top quark and Higgs boson decays in Alternative Left-Right 
Symmetric Models}

\author{R. Gait\'an$^1$}
 \email[Ricardo.Gaitan@fis.cinvestav.mx]{}
\author{O. G. Miranda$^2$}
 \email[omr@fis.cinvestav.mx]{}
\author{L. G. Cabral-Rosetti$^3$}
 \email[luis@nuclecu.unam.mx]{}

\affiliation{$^1$Centro de Investigaciones Te\'oricas,\\
                 Facultad de Estudios Superiores -- Cuautitl\'an,\\
                 Universidad Nacional Aut\'onoma de M\'exico, (FESC-UNAM).\\
                 A. Postal 142,Cuautitl\'an-Izcalli, Estado de M\'exico, 
                 C. P. 54700, M\'exico.\\
\\             $^2$Departamento de F{\'\i}sica,\\
                 Centro de Investigaci\'on y de Estudios Avanzados del IPN\\
                 A. Postal 14-740, M\'exico D. F. 07000, M\'exico.\\
\\             $^3$Instituto de Ciencias Nucleares,\\
                 Departamento de F{\'\i}sica de Altas Energ{\'\i}as,\\
                 Universidad Nacional Aut\'onoma de M\'exico, (ICN-UNAM).\\
                 Circuito Exterior, C.U., A. Postal 70-543, 
                 04510 M\'exico, D.F.,  M\'exico.}

\begin{abstract} 
Top quark and Higgs boson decays induced by flavor-changing neutral 
currents (FCNC) are very much suppressed in the Standard Model (SM). 
Their detection in colliders like the Large Hadron Collider (LHC),
Next Linear Collider (NLC) or Tevatron would be a signal of new
physics.
We evaluate the FCNC decays $t \rightarrow H^0 + c$, $t \rightarrow Z
+ c$, and $H^0 \rightarrow t + {\bar c}$ in the context of Alternative
Left-Right symmetric Models (ALRM) with extra isosinglet heavy
fermions; in this case, FCNC decays occurs at tree-level and they are
only suppressed by the mixing between ordinary top and charm quarks,
which is poorly constraint by current experimental values.
This provides the possibility for future colliders, either to detect
new physics, or to improve present bounds on the parameters of
the model. 
\end{abstract}

\maketitle

%%%%%%%%%%%%%%%%%%%%%%%%%%%%%%%%%%%%%%%%%%%%%%%%%%%%%%%%%%%%%%%%%%%%%%%
\section{Introduction}
%%%%%%%%%%%%%%%%%%%%%%%%%%%%%%%%%%%%%%%%%%%%%%%%%%%%%%%%%%%%%%%%%%%%%%%

Rare top quark decays are interesting because they might be a source
of possible new physics effects. 
Due to its large mass of about $178\, GeV$ \cite{Abazov:2004cs}, the
top quark dominant decay mode is into the channel $t \rightarrow b +
W$.
In the Standard Model (SM),
based on the spontaneously broken local symmetry $SU(3)_C \otimes
SU(2)_L\otimes U(1)_Y$, flavor-changing neutral currents (FCNC) are
absent at the tree-level due to the Glashow-Iliopoulos-Maiani (GIM)
mechanism, and they are extremely small at loop level. However, new FCNC
states can appear in top decays if there is physics beyond the
Standard Model.  Moreover, in some particular models
beyond the SM, rare top decays may be significantly enhanced to reach
detectable levels \cite{Bejar1}.

Rare top decays have been studied in the context of the SM and beyond
\cite{Jenkins,Mele,Diaz:91}. The top quark decays into gauge bosons
($t \rightarrow c + V;\, V \equiv \gamma,\, Z,\, g$) are extremely
rare events in the SM; their branching ratios are, according to Ref.
\cite{Diaz:91,Eilam}: $5 \times 10^{- 13}$ for the photon, $10^{- 13}$
for the Z-boson and $\sim 4 \times 10^{- 11}$ for the gluon channel,
and even smaller according to other
estimates~\cite{Aguilar}. Similarly, the top quark decay into the SM
Higgs boson, is a very rare decay, with $BR\, (t \rightarrow c + H)
\sim 10^{- 14}$ \cite{Mele,EilamE}.  However, by considering physics
beyond the SM, for example, the Minimal Supersymmetric Standard Model
(MSSM) or the two-Higgs-doublet model (2HDM) or extra quark singlets,
new possibilities open up \cite{Bejar1, Jenkins, Mele, Han, Eilam,
Aguilar, EilamE, Herrero, Aguilar-Saavedra:2002kr}, enhancing this
branching ratios to the order of $\sim 10^{-6}$ for the $t\to c +
Z$~\cite{Aguilar} channel and $\sim10^{-4}$ for the $t\to c+
H$~\cite{Herrero} case. The rare top decay $t \rightarrow q+W+Z$ has
also been considered as a future test of new physics~\cite{Diaz}.

On the other hand, the FCNC decays of the Higgs boson can be
important in various scenarios, including the MSSM \cite{Hou}. The FCNC
Higgs decay into a top quark within a general 2HDM has been studied
in Ref.  \cite{Bejar2}. 
Because the FCNC Higgs decays in the SM are very suppressed, any
experimental signature of Higgs FCNC type could be evidence of physics
beyond the SM.

In the future CERN Large Hadron Collider (LHC), about $10^7$ top quark
pairs will be produced per year~\cite{Beneke}. An eventual signal of
FCNC in the top quark decay will have to be ascribed to new
physics. Furthermore, since the Higgs boson could also be produced at
significant rates in future colliders, it is also important to search
for all the relevant FCNC Higgs decays.

On the other hand, while the electroweak SM has been successful in the
description of low-energy phenomena, it leaves many questions
unanswered. One of them has to do with the understanding of the origin
of parity violation in low-energy weak interaction processes. Within
the framework of left-right symmetric models, based on the gauge group
$SU(2)_L \otimes SU(2)_R \otimes U(1)_{B-L}$, this problem finds a
natural answer\cite{pati-salam,Mohapatra-book}. Moreover, new
formulations of this model have been considered in which the fermion
sector has been enlarged to include isosinglet vectorlike heavy
fermions in order to explain the mass
hierarchy~\cite{Davidson-Wali,Kiers}, the smallness of the neutrino
mass~\cite{Babu-Xe} or the problem of weak and strong
CP-violation~\cite{Balakrishna,barr}. Most of these models includes
two Higgs doublets.

In this paper we consider the rare top decay into a Higgs boson and
the FCNC decay of the Higgs boson with the presence of a top quark in
the final state, within the context of these alternative left-right
models (ALRM) with extra isosinglet heavy fermions. Due to the
presence of extra quarks the Cabibbo-Kobayashi-Maskawa matrix is not
unitary and FCNC may exist at tree-level.

Therefore, a high branching ratio for the decay $t \rightarrow c + H$
(for a Higgs boson lighter than the top quark mass) or for the decay
$H\rightarrow t + \bar{c}$ is allowed, opening great opportunities
either to detect or to constrain the mixing parameter $\eta_{\chic {32}}$
between the ordinary top and charm quarks.

The organization of the paper is as follows: in Section 2 we review
the alternative left-right model (ALRM) giving emphasis to the fermion
mixing and flavor violation. In Section 3 we present our calculations in the
ALRM for the processes $t\rightarrow c +Z$, $t \rightarrow H^0 + c$
and $H^0 \rightarrow t + {\bar c}$; we derive bounds on the parameters
of the model associated with FCNC transitions and we discuss future
perspectives for improving this bounds. Section 4 contains our
conclusions.
 
%%%%%%%%%%%%%%%%%%%%%%%%%%%%%%%%%%%%%%%%%%%%%%%%%%%%%%%%%%%%%%%%%%%%%%%%%%
\section{The Model}
%%%%%%%%%%%%%%%%%%%%%%%%%%%%%%%%%%%%%%%%%%%%%%%%%%%%%%%%%%%%%%%%%%%%%%%%%%%

The ALRM formulation is based on the gauge group $SU(2)_{\chic L}
\otimes SU(2)_{\chic R} \otimes U(1)_{B-L}$. In order to solve
different problems such as the hierarchy of quark and lepton masses or
the strong CP problem, different authors have enlarged the fermion
content to be of the form

\begin{eqnarray}
l^{\chic {0}}_{i\, {\chic {L}}} =
     \left( \begin{array}{c}
     \nu^{\chic {0}}_i \\ {e^{\chic {0}}_i}\end{array} \right)_{\chic {L}}
   ,\ {e}^{\chic {0}}_{i\,{\chic {R}}}\ \ \ \ \ &;& \ \ \ \ \
 {\widehat l}^{\chic {0}}_{i\, {\chic {R}}} = 
     \left( \begin{array}{c}
     {\widehat {\nu}^{\chic {0}}_{i}} \\
     {{\widehat {e}}^{\chic {0}}_{i}} \end{array} \right)_{\chic {R}}
   ,\ {\widehat {e}}^{\chic {0}}_{i\,{\chic {L}}}  \nonumber \\ 
Q^{\chic {0}}_{i\, {\chic {L}}} =
     \left( \begin{array}{c}
     {{u}^{\chic {0}}_{i}}\\
     {{d}^{\chic {0}}_{i}}
     \end{array} \right)_{\chic {L}}
   ,\ {u}^{\chic {0}}_{i\,{\chic {R}}}\
   ,\ {d}^{\chic {0}}_{i\,{\chic {R}}}, \ \ \ \ \ &;& \ \ \ \ \
{\widehat Q}^{\chic {0}}_{i\, {\chic {R}}} =
     \left( \begin{array}{c}
          {\widehat {u}^{\chic {0}}_{i}}\\
     {{\widehat {d}}^{\chic {0}}_{i}}
     \end{array} \right)_{\chic {R}}
   ,\ {\widehat {u}}^{\chic {0}}_{i\,{\chic {L}}}\
   ,\ {\widehat {d}}^{\chic {0}}_{i\,{\chic {L}}}\ ,
\label{Eq2.1}
\end{eqnarray}

where the index $i$ ranges over the three fermion families.
The superscript $0$ denote weak eigenstates. The quantum numbers of 
these fermions, under the gauge group 
$SU(3)_C\otimes SU(2)_L \otimes SU(2)_R \otimes U(1)_{B-L}$, are given by 

\begin{eqnarray}
l^{\chic {0}}_{i\, {\chic {L}}} \quad (1,2,1)_{-1}
\quad
{e}^{\chic {0}}_{i\,{\chic {R}}}\quad (1,1,1)_{-2}&;&
\qquad
{\widehat l}^{\chic {0}}_{i\, {\chic {R}}} \quad (1,1,2)_{-1}
\quad
{\widehat {e}}^{\chic {0}}_{i\,{\chic {L}}} \quad (1,1,1)_{-2} \nonumber \\
{u}^{\chic {0}}_{i\,{\chic {R}}} \quad (3,1,1)_{\frac43}
\qquad
{d}^{\chic {0}}_{i\,{\chic {R}}} \quad (3,1,1)_{\frac23}&;&
\qquad
{\widehat{u}}^{\chic {0}}_{i\,{\chic {L}}} \quad (3,1,1)_{\frac43}
\qquad
{\widehat {d}}^{\chic {0}}_{i\,{\chic {L}}} \quad (3,1,1)_{\frac23} 
\nonumber\\
Q^{\chic {0}}_{i\, {\chic {L}}} \quad (3,2,1)_{\frac13}
& &\qquad
{\widehat{Q}}^{\chic {0}}_{i\, {\chic {R}}} \quad (3,1,2)_{\frac13}
\label{Eq2.1.b}
\end{eqnarray}

In many of these models, extra neutral leptons also appears in order
to explain the neutrino mass pattern, however we will focus in this
work only on the quark sector. 

In order to break $SU(2)_{\chic L} \otimes SU(2)_{\chic R} \otimes
  U(1)_{B-L}$ down to $U(1)_{\chic {em}}$ the ALRM introduces
  two Higgs doublets. The SM one ($\phi$) and its partner (${\widehat
  \phi}$). The symmetry breaking is done in such a way that the
  vacuum expectation values of the Higgs fields are

\be
\big <  \phi \big > = \frac{1}{\sqrt{2}}
\left(\begin{array}{c}
{0} \\
{v}
\end{array}\right)\ \ \ \ \ ;\ \ \ \ \
\big <  {\widehat {\phi}} \big > = \frac{1}{\sqrt{2}}
\left(\begin{array}{c}
{0} \\
{\widehat {v}}
\end{array}\right)\ .
\label{Eq2.2}
\ee

Ref. \cite{Ceron} shows that from the eight scalar degrees of freedom,
six become the Goldstone bosons required to give mass to the $W^\pm$,
${\widehat W}^\pm$, $Z$ and ${\widehat Z}$; thus two neutral Higgs
bosons remain in the physical spectrum. The neutral physical states
are

\be
H = {\sqrt 2} \Big( \Re e\, {\phi}^0 - v \Big)\, \cos \alpha
+ \Big( \Re e\, {\widehat {\phi}}^0 - {\hat v} \Big)\, \sin\, \alpha
\label{Eq2.3}
\ee

\be
{\widehat H} = - {\sqrt 2} \Big( \Re e\, {\phi}^0 - v \Big)\,
\sin \alpha + \Big( \Re e\, {\widehat {\phi}}^0 - {\hat v} \Big)\,
\cos \alpha\ ,
\label{Eq2.4}
\ee where $\alpha$ denotes the neutral Higgs mixing angle (which
diagonalize the neutral Higgs mass matrix). The renormalizable and
gauge invariant interactions of the scalar doublets $\phi$ and
${\widehat \phi}$ with the fermions are described by the Yukawa
Lagrangian. For the quark fields, the corresponding Yukawa terms are
written as

\be
\ba{c}
\displaystyle
{\cal L}_{\chic {Y}}^{q} = {\lam}^{d}_{i j}\,
{\overline {Q^{\, {\chic 0}}_{i {\chic L}}}}\,
{\phi}\, d^{\, {\chic 0}}_{j {\chic R}}
+ {\lam}^{u}_{i j}\, {\overline {Q^{\, {\chic 0}}_{i {\chic L}}}}\,
{\widetilde {\phi}}\, u^{\, {\chic 0}}_{j {\chic R}}
+ {\widehat \lam}^{d}_{i j}\,
{\overline {\widehat Q^{\, {\chic 0}}_{i {\chic R}}}}\,
{\widehat \phi}\, {\widehat d}^{\, {\chic 0}}_{j {\chic L}}
\\[0.3cm]
\displaystyle
+ {\widehat \lam}^{u}_{i j}\,
{\overline {\widehat Q^{\, {\chic 0}}_{i {\chic R}}}}\,
{\widetilde {\widehat {\phi}}}\, {\widehat u}^{\, {\chic 0}}_{j {\chic L}}
+ {\mu}^{d}_{i j}\, {\overline {\widehat d^{\, {\chic 0}}_{i {\chic L}}}}\,
d^{\, {\chic 0}}_{j {\chic R}}
+ {\mu}^{u}_{i j}\, {\overline {\widehat u^{\, {\chic 0}}_{i {\chic L}}}}\,
u^{\, {\chic 0}}_{j {\chic R}} + h. c.
\ea
\label{Eq2.5}
\ee where $i,j = 1, 2, 3$ and ${\lam}^{d (u)}_{i j}$, ${\widehat
  \lam}^{d (u)}_{i j}$, and ${\mu}^{d (u)}_{i j}$ are (unknown)
matrices. The conjugate fields ${\widetilde {\phi}}$ $\Big(
{\widetilde {\widehat {\phi}}} \Big)$ are ${\widetilde {\phi}} = i
\tau_2 \phi^*$ and ${\widetilde {\widehat {\phi}}} = i \tau_2
{\widehat \phi^*}$, with $\tau_2$ the Pauli matrix. 

We can introduce the generic vectors~\cite{Langacker:1988ur}
$\psi^0_L$ and $\psi^0_R$ , for representing left and right
electroweak states with the same charge. These vectors can be
decomposed into the ordinary and exotic sector by

\be
\psi^0_L = 
\left(\begin{array}{c}
\psi^0_{OL} \\ \\ \psi^0_{EL}  \end{array}\right) 
\qquad \psi^0_R = 
\left(\begin{array}{c}
\psi^0_{OR} \\ \\ \psi^0_{ER}  \end{array}\right),
\ee 
where $\psi^0_{OL}$ is a column vector consisting of the SM $SU(2)_L$
doublets (for example the $u^{\chic {0}}_{i\, {\chic {L}}}$) while
$\psi^0_{EL}$ contains the exotic singlets ( ${\widehat {u}}^{\chic
{0}}_{i\,{\chic {L}}}$). The vector $\psi^0_{OR}$ contains the SM
singlets (like $u^{\chic {0}}_{i\, {\chic {R}}}$) and $\psi^0_{ER}$
contains the exotic $SU(2)_R$ doublets (${\widehat {u}}^{\chic
{0}}_{i\,{\chic {R}}}$).

In the same way we can define the vectors for the mass eigenstates in
terms of 'light' and 'heavy' states: 

\be \psi_L =
\left(\begin{array}{c}
\psi_{lL} \\ \\ \psi_{hL}  
\end{array}\right) 
\qquad \psi_R = 
\left(\begin{array}{c}
\psi_{lR} \\ \\ \psi_{hR}  
\end{array}\right) 
\ee 
The relation between weak eigenstates and mass eigenstates will be 
given through the matrices $U_L$ and $U_R$: 

\be 
\psi^0_L = U_L \psi_L \qquad \psi^0_R =U_R \psi_R 
\ee
where 

\be
{\sf U}_a = \left(\begin{array}{ccc}
{{\sf A}_a} &  & {{\sf E}_a}  \\
{{\sf F}_a} &  & {{\sf G}_a}
\end{array}\right) 
, \qquad a = L,R
\label{Eq2.7}
\ee
Here, $A_a$ is the $3\times 3$ matrix relating the ordinary weak
states with the light-mass eigenstates, $G_a$ is a $3\times3$ matrix
relating the exotic states with the heavy ones, while $E_a$ and $F_a$
describe the mixing between the two sectors. 

It is easy to see that in this case, the $A_a$ is not necessarily
unitary. Instead the unitarity of the $U_a$ matrices leads to the
relations 

\begin{eqnarray}
A^\dagger_a A_a + F^\dagger_a F_a &=& I \nonumber \\
A^\dagger_a A_a + E^\dagger_a E_a &=& I .
\end{eqnarray}

Therefore, in this model, thanks to the extra heavy quarks,  it is
possible to have a relatively big mixing between ordinary quarks. This 
is not a particular characteristic of the model but a general feature when 
considering models with extra heavy singlets~\cite{Aguilar-Saavedra:2004wm}.

The tree-level interaction of the neutral Higgs bosons $H$ and
${\widehat H}$ with the light fermions are given by

\be
\ba{c}
\displaystyle
{\cal L}_{\chic {Y}}^{f} = \frac{g}{2 \sqrt 2}
{\overline \psi_{\chic L}} A^{\dag}_{\chic L} A_{\chic L}
\frac{m_f}{M_{\chic W}} \psi_{\chic R} \Big( H\, \cos \alpha
- {\widehat H}\, \sin \alpha \Big)
\\[0.3cm]
+ \displaystyle
\frac{{\widehat g}}{\sqrt 2}
{\overline \psi_{\chic L}} \frac{m_f}{M_{\chic {\widehat W}}}
F^{\dag}_{\chic R} F_{\chic R} \psi_{\chic R}\Big( H\, \sin \alpha
+ {\widehat H}\, \cos \alpha \Big) + h. c.
\ea
\label{Eq2.6}
\ee 

The neutral current in terms of the mass eigenstates, including the
contribution of the neutral gauge boson mixing, can be written as
follows:

\begin{equation}
- {\cal L}^{\, n.\, c.} = 
\sum_{a= {\chic L}\, {\chic R}}\;
{\overline \psi_{a}}\, \gamma^{\mu}\, {\sf U}_a^{\dag}\,
\Big( g\, {\sf T}_{3a},\, {\widehat g}\, {\widehat {\sf T}}_{3a},\,
g^{'}\, \frac{Y_{a}}{2} \Big)\, {\sf U}_a\, \psi_{a}\,
\left(\begin{array}{c}
{Z} \\
{{\widehat Z}} \\
{A}
\end{array}\right)
\label{Eq2.9}
\end{equation}
where ${\sf T}_{3a}$, ${\widehat {\sf T}}_{3a}$, and $Y$ are the
generators of the $SU(2)_{\chic L}$, $SU(2)_{\chic R}$, and
$U(1)_{B-L}$, respectively. For the sake of simplicity, we 
will consider only the case $g = \widehat g$.

From the last two equations we can see that, thanks to the
non-unitarity of the $A_{a}$ matrices we can have FCNC at tree-level. 
This characteristic appears due to the extra quark content
of the model, which is not present in the usual left-right symmetric
model.

%%%%%%%%%%%%%%%%%%%%%%%%%%%%%%%%%%%%%%%%%%%%%%%%%%%%%%%%%%%%%%%%%%%%%%%%%%%
\section{FCNC top and Higgs decays in the ALRM}
%%%%%%%%%%%%%%%%%%%%%%%%%%%%%%%%%%%%%%%%%%%%%%%%%%%%%%%%%%%%%%%%%%%%%%%%%%%

Once we have introduced the model we are interested in, we compute the
expected branching ratio for a FCNC top or Higgs decay with a charm
quark in the final state. We perform this analysis in this section. We
will start by searching the maximum allowed value for a top-charm
mixing and then we will obtain the possible branching ratio both for
the top decay into a Higgs boson plus a charm quark and for the Higgs
decay into a top plus an anti-charm quark.

%%%%%%%%%%%%%%%%%%%%%%%%%%%%%%%%%%%%%%%%%%%%%%%%%%%%%%%%%%%%%%%%%%%%%
\subsection{Constraining the top-charm mixing angle}
%%%%%%%%%%%%%%%%%%%%%%%%%%%%%%%%%%%%%%%%%%%%%%%%%%%%%%%%%%%%%%%%%%%%%%

In order to have an expectation on the branching ratio for the FCNC
top decay in the ALRM we need first an estimate on the mixing between
the top and charm quarks in the model. One may think that the best
constrain could come from the flavor-changing coupling of the
neutral $Z$ boson to the top and charm quarks, which can be written
as:

\be
\ba{c}
\displaystyle
{\cal L}_{Z}^{c t} = \frac{e}{s_{\theta_{\chic W}}\, c_{\theta_{\chic W}}}\,
\overline c\, (g_{\chic V} + g_{\chic A})\, \gamma^{\mu}\, Z_{\mu}\, t
\ea
\label{eq:Lag.tzc}
\ee
where

\be \ba{c} g_{\chic {V}, {\chic A}} = \frac{1}{4}\, (c_{\Theta} -
\frac{s^{2}_{\theta_{\chic W}}}{r_{\theta_{\chic W}}}\, s_{\Theta})\, 
\eta^{L}_{\chic {32}} \pm 
\frac{1}{4}\, \frac{c^{2}_{\theta_{\chic W}}}{r_{\theta_{\chic W}}}\,
s_{\Theta}\, \eta^{R}_{\chic {32}} \ea \ee 
and $s_{\theta_{\chic W}}$, $c_{\theta_{\chic W}}$ and
$r_{\theta_{\chic W}}$ are, respectively, $\sin\theta_{\chic W}$,
$\cos\theta_{\chic W}$ and $\sqrt{\cos^{2}{\theta_{\chic W}} -
\sin^{2}{\theta_{\chic W}}}$; $\theta_{\chic W}$ is the weak mixing
angle, $\Theta$ is the mixing between the $Z$ and $\widehat{Z}$
neutral gauge bosons.
Here, $\eta^L_{\chic {32}}$ and $\eta^{R}_{\chic {32}}$ represent the mixing
between the ordinary top and charm quarks and are given by
\be \eta^L_{\chic {32}} = (A^{+}_{L}\, A_{L})_{\chic
    {32}}\qquad 
\eta^R_{\chic {32}} 
= (A^{+}_{R}\, A_{R})_{\chic
    {32}}.  \ee 
Since the mixing between the $Z$ and the $\widehat{Z}$ neutral gauge
bosons, $\Theta$, is expected to be small~\cite{Adriani} it can be safely
neglected, and this partial width will not depend on the parameter 
$\eta^R_{\chic {32}}$. Therefore, from now on we will denote 
$\eta_{\chic {32}} = \eta^L_{\chic {32}}$.

From Eq. (\ref{eq:Lag.tzc}) we can compute the branching ratio for the
decay $t \to Z +c$ and compare it to the experimental limit $B(t
\rightarrow Z + c) \leq 0.137$~\cite{Abbiendi} at 95 \% C. L. We will
get the maximum value for $\eta_{\chic {32}} \leq 0.53$.

Although we have found a direct constrain to $\eta_{\chic {32}}$, it
is possible to get a stronger limit if we use the unitarity properties
of the mixing matrix and the constrain on $\eta_{\chic {22}}$ that
comes from the branching ratio $\Gamma (Z \rightarrow c + \bar{c})$.
The experimental value for the branching ratio of this process is
given by $B(Z\to c\bar{c})=\Gamma(Z\to c\bar{c})/\Gamma_{\chic{
total}}=0.1181 \pm 0.0033$ (see~\cite{pdg}). Using this experimental
value, the minimum value for $\eta_{\chic {22}}$ at 95 \% C. L. will be
$\eta_{\chic {22}} \ge 0.99$.

This information is of great help for constraining $\eta_{\chic{32}}$
since the unitarity of the mixing matrix has already been analyzed in
the general case~\cite{Aguila-PRL} and leads to the following relation:

\be
|\eta_{\chic{32}}|^2 \le (1 - \eta_{\chic{33}})(1 - \eta_{\chic{22}}).
\label{eq:unitary}
\ee

Although we don't know the value for $\eta_{\chic{33}}$, the boundary
on $\eta_{\chic{22}}$ is enough to see that the mixing parameter
$\eta_{\chic{32}} \le 0.1$. The higher value $\eta_{\chic{23}} = 0.1$
is obtained when we take the extreme case $\eta_{\chic{33}} = 0$, as
can be seen from Eq. (\ref{eq:unitary}). 

It is possible to obtain more stringent constraints if low-energy data
are considered. For the case of two extra quark singlets, this
analysis was done in a very general framework in
Ref. \cite{Aguilar-Saavedra:2002kr}. After a very complete analysis of
all the observables, the author of this article obtained 
$|\eta_{\chic{32}}|\leq 0.036$. This relatively large value is allowed
for the case of a exotic top mass similar to that of the SM
top-quark~\footnote{There are not stringent lower bounds on the mass
of a exotic top quark, being $220$~GeV the current direct
limit~\cite{Acosta03}}. In the case of a very heavy mass for the
exotic top-quark the constraint is more stringent:
$|\eta_{\chic{32}}|\leq 0.009$. In what follows we will use these two
values in order to illustrate the expected signals from rare Higgs and
top decays.

%%%%%%%%%%%%%%%%%%%%%%%%%%%%%%%%%%%%%%%%%%%%%%%%%%%%%%%%%%%%%%%%%%%%%%%%%
\subsection{The decay $t \rightarrow H^0 + c$}
%%%%%%%%%%%%%%%%%%%%%%%%%%%%%%%%%%%%%%%%%%%%%%%%%%%%%%%%%%%%%%%%%%%%%%%%%

Now that we have an estimate for the value of $\eta_{\chic {32}}$, we 
compute the branching ratio for $t \rightarrow H^0 + c$ in the
framework of ALRM. We take the charged-current two-body decay $t
\rightarrow b + W$ to be the dominant t-quark decay mode. The neutral
Higgs boson $H^0$ will be assumed to be the lightest neutral mass
eigenstate.
Assuming $M_{\widehat{H}} \gg M_H$ the vertex $tcH^{0}$ is written as
follows:

\be
\frac{g\, m_t\, \eta_{\chic {32}}}{2 M_{\chic {W}}}\, \cos \alpha \, P_L
\label{luis}
\ee
The partial width for this tree-level process can be obtained 
in the usual way and it is given by:

\be
\ba{c}
\displaystyle
\Gamma(t \rightarrow H^0 + c) =\\[0.5cm]
\displaystyle
  \frac{G_{\chic F}\, \eta^2_{\chic {32}}\, \cos^2
\alpha} {16\, \sqrt{2}\, \pi\, m_t}\, \Big( m_t^2 +  m_c^2 -
M_{\chic H}^2 \Big ) \Big[ \Big( m_t^2 -  \Big( M_{\chic H} + m_c
\Big )^2\, \Big) \Big( m_t^2 -  \Big( M_{\chic H} - m_c  \Big)^2\, 
\Big) \Big]^{\frac{1}{2}}
\ea
\label{ricard}
\ee 
where $G_{\chic F}$ is the Fermi's constant, $m_t$ denotes the top
mass, $m_c$ is the charm mass, and $M_{\chic H}$ is the mass of the
neutral Higgs boson.  We can see from this formula that the branching
ratio will be proportional to the product $\eta_{\chic {32}}\cos
\alpha$, of the top-quark mixing with the SM Higgs boson mixing with
the extra Higgs boson.

\begin{figure*}
\includegraphics[width=0.8\textwidth,angle=-90]{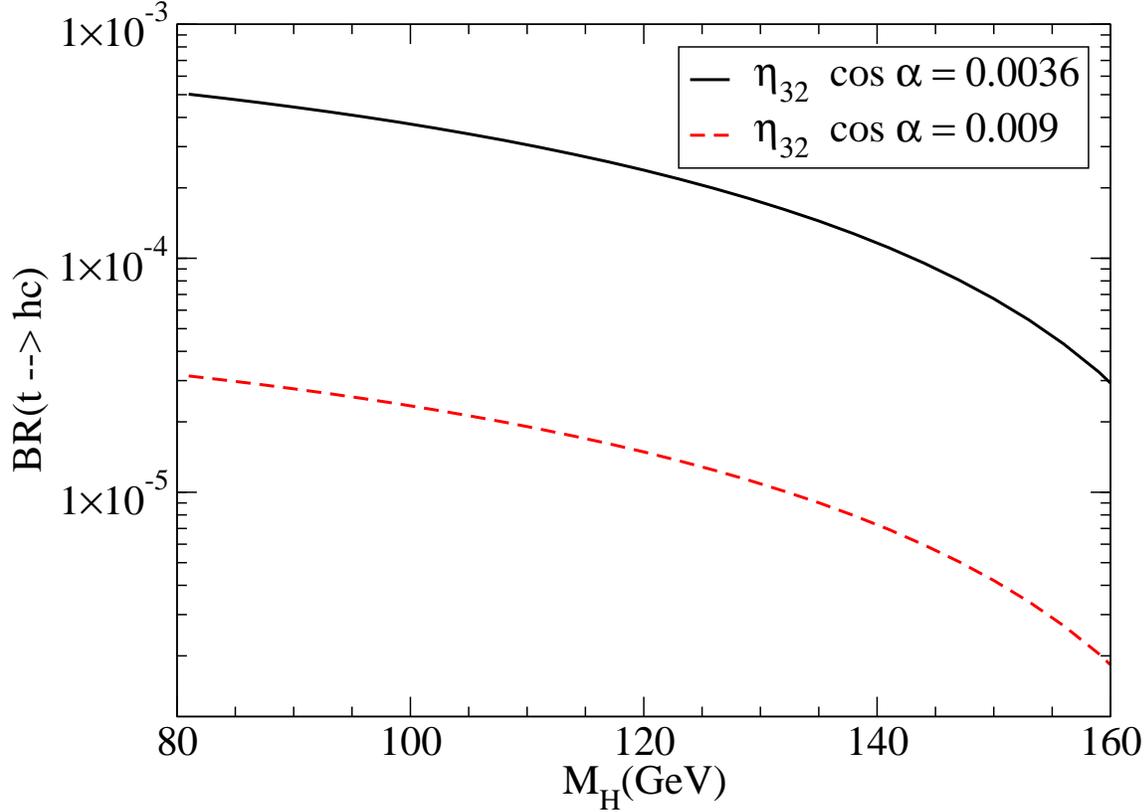}
       \caption{
	 Branching ratio for the rare top decay $t\rightarrow H + c$ for
  different values of the product of the mixing, $\alpha$, between the
  lightest Higgs bosons and the additional Higgs boson of the model,
  and the mixing between the top quark and the charm quark,
  $\eta_{\chic {32}}$. This figure shows that there is a lot of room
  for future collider experiments either to detect, or to set bounds
  on this parameters.
	\label{fig:top-decay}}
\end{figure*}

The branching ratio for this decay is obtained as the ratio of Eq. 
(\ref{ricard}) to the total width for the top quark, namely 

\be
\ba{c}
B(t \rightarrow H^{0} + c)=
\frac{\Gamma(t \rightarrow H^{0} + c)}{\Gamma(t \rightarrow b + W)} .
\ea
\ee

Thanks to the possible combined effect of a big $\cos\alpha$ (null
mixing between the SM Higgs boson and the additional Higgs bosons) and
a big value of $\eta_{\chic {32}}$ this branching ratio could be as
high as $\approx 3\times 10^{-4}$, for a Higgs mass of 117~$GeV$ as is
illustrated in Fig.  \ref{fig:top-decay}. Perhaps is more realistic to
consider the more stringent constraint $\eta_{\chic {32}} = 0.009$,
but even in this case, for $\cos \alpha \approx 1$ there is still
sensitivity for detecting a positive signal of order $10^{-5}$ as is
shown in same Fig.~\ref{fig:top-decay}.

%%%%%%%%%%%%%%%%%%%%%%%%%%%%%%%%%%%%%%%%%%%%%%%%%%%%%%%%%%%%%%%%%%%%%%%%%%%
\subsection{The decay $H^0 \rightarrow t + \bar{c}$}
%%%%%%%%%%%%%%%%%%%%%%%%%%%%%%%%%%%%%%%%%%%%%%%%%%%%%%%%%%%%%%%%%%%%%%%%%%%

Finally we also consider the case of a Standard Higgs with a large
mass. The best-fit value of the expected Higgs mass, including the new
average for the mass of the top quark, is 117 GeV~\cite{Abazov:2004cs}
and the upper bound is $M_H \le 251$~GeV at 95 \% C L. However, the
error for the Higgs boson mass from this global fit is asymmetric, and
a Higgs mass of $400$~GeV is well inside the $3\sigma$ region as can
be seen in Ref~\cite{Abazov:2004cs}.

We estimate the branching ratio for the decay $H^0 \rightarrow t +
\bar{c}$, where $H^{0}$ is the light neutral Higgs boson of the
ALRM. The expression for the partial width is

\be
\ba{c}
\displaystyle
\Gamma(H^0 \rightarrow t + \bar{c}) =\\[0.5cm]
\displaystyle 
 \frac{3\, G_{\chic F}\, m_t^2\, \eta^2_{\chic
{32}}\, \cos^2 \alpha} {8\, \sqrt{2}\, \pi\, M_{\chic H}^3}\,
\Big( M_{\chic H}^2 - m_t^2 -  m_c^2 \Big ) \Big[ \Big( M_{\chic H}^2 
-  \Big( m_t + m_c \Big )^2\, \Big) \Big( M_{\chic H}^2 -
\Big( m_c - m_t \Big )^2\, \Big) \Big]^{\frac{1}{2}}.
\ea
\label{Eq2.11}
\ee

The branching ratio for this decay is obtained as the ratio of Eq.
(\ref{Eq2.11}) to the total width of the Higgs boson, which will
include the dominant modes $H^{0} \rightarrow b + \bar{b}$, $H^{0}
\rightarrow c + \bar{c}$, $H^{0} \rightarrow \tau + \bar{\tau}$,
$H^{0} \rightarrow W + W$, and $H^{0} \rightarrow Z + Z$. The
expressions for these decay widths in the ALRM are:

\be
\ba{c}
\displaystyle
\Gamma(H^{0} \rightarrow f + \bar{f}) =
C_f\, \frac{G_F\, m^{2}_f\, M_H\, \eta^{2}_{\chic {ff}}\, \cos^{2}
\alpha}{4 \sqrt{2}\, \pi} (1 - 4\lambda_f)^{\frac{3}{2}}
\ea
\ee
where $\lambda_f= (\frac{m_{f}}{M_H})^{2}$, 
$C_f$ = 1 for leptons and $C_f$ = 3 for quarks,

\be
\ba{c}
\displaystyle
\Gamma(H^{0} \rightarrow W + W) =
\frac{G_F\, M_H^{3}\, \cos^{2}\alpha}{8 \sqrt{2}\, \pi} (1 - 4
\lambda_{W})^{\frac{1}{2}} (1 - 4 \lambda_{W} + 12
\lambda_{W}^{2}),
\ea
\ee
with $\lambda_{W}$ = $(\frac{M_W}{M_H})^{2}$, and

\be
\ba{c}
\displaystyle
\Gamma(H^{0} \rightarrow Z + Z) =
\frac{G_F\, M_H^{3}\, M_W^{4}\, X^{2}\, \cos^{2}\alpha}{16
\sqrt{2}\, \pi\, M^4_Z} (1 - 4 \lambda_{Z})^{\frac{1}{2}} (1 - 4
\lambda_{Z} + 12 \lambda_{Z}^{2}),
\ea
\ee
with $\lambda_{Z}$ = $(\frac{M_Z}{M_H})^{2}$ and $X$ = 
$(c_{\theta_{\chic W}}c_{\Theta} -
\frac{s_{\theta_{\chic W}}}{r_{\theta_{\chic W}}} 
t_{\theta_{\chic W}}(s_{\Theta} - r_{\theta_{\chic W}}c_{\Theta}))^{2}$.

\begin{figure*}
\includegraphics[clip,width=0.5\textwidth,angle=-90]{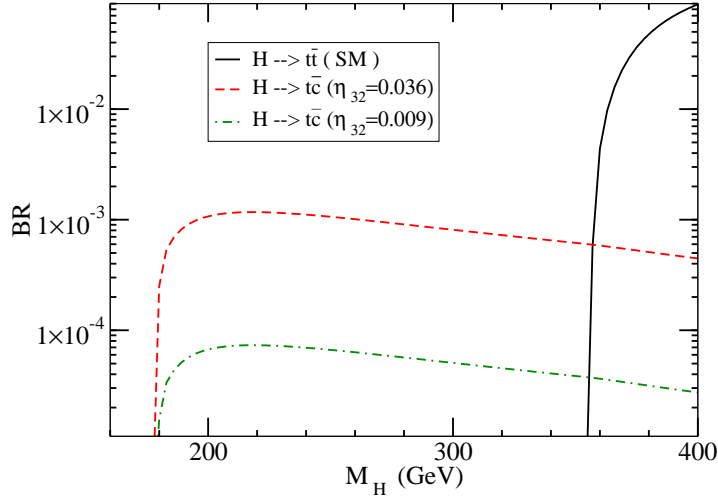}
\caption{\label{fig:higgs-decay}
  Branching ratio for the  rare Higgs decay $H\rightarrow
  t + \bar{c}$, for different values of $\eta_{\chic {32}}$ as a
  function of the Higgs mass. The standard Higgs decay $H\rightarrow t
  + \bar{t}$ is also shown.
}
\end{figure*}

We show in Fig. \ref{fig:higgs-decay} the branching ratios for
different decay modes, both for the Standard Model case ($\eta_{\chic {32}} =
0$ and $\eta_{ii}=1$) and for the FCNC case. We can see that, also for
a heavy Higgs, there are chances to either detect or to constrain the
mixing angle parameter $\eta_{\chic {32}}$.  In this case, since all the
partial widths have the same dependence on $\cos^{2} \alpha$, the
branching ratios will depend only on $\eta_{\chic {32}}$.

%%%%%%%%%%%%%%%%%%%%%%%%%%%%%%%%%%%%%%%%%%%%%%%%%%%%%%%%%%%%%%%%%%%%%%%%%%%
\section{Results and conclusions}

We have seen that the ALRM allows relatively big values of
$\eta_{\chic {32}}$. The $t \to H + c$ branching ratio could be of
order of $10^{-4}$, which is at the reach of LHC.
For example, it has been estimated that the LHC sensitivity (at 95 \%
C. L.)  for this decay  is $Br (t\to Hc)\leq4.5\times
10^{-5}$ \cite{Aguilar-Saavedra:2000aj};
this branching ratio would be obtained in this model for a top-charm
mixing $\eta_{\chic {32}} = 0.015$ and a diagonal ordinary top
coupling $\eta_{\chic {22}} \simeq 0.98$.
On the other hand, the FCNC mode $H \to t + \bar{c}$ may reach a
branching ratio of order $10^{-3} $ and can also be a useful
channel to look for signals of physics beyond the SM in the LHC.

The ALRM is a well motivated model that rises from different
theoretical motivations and has a rich phenomenology.  In particular,
we have studied the ALRM in the context of rare top decays and we have
found that these models could be tested in the next generation of
colliders.

%%%%%%%%%%%%%%%%%%%%%%%%%%%%%%%%%%%%%%%%%%%%%%%%%%%%%%%%%%%%%%%%%%%%%%%%%%%%

\acknowledgments We would like to thanks Lorenzo Diaz Cruz for
  bringing to our attention the subject of rare top decays.
  R. G. L. would like to thanks CINVESTAV-IPN for the nice environment
  during his sabbatical period in this institution. This work has been
  supported by Conacyt and SNI.

%%%%%%%%%%%%%%%%%%%%%%%%%%%%%%%%%%%%%%%%%%%%%%%%%%%%%%%%%%%%%%%%%%%%%%%%%%%%
%

%
%%%%%%%%%%%%%%%%%%%%%%%%%%%%%%%%%%%%%%%%%%%%%%%%%%%%%%%%%%%%%%%%%%%%%%%%%%%%

\end{document}